# Magneto-Optical Modulation Ellipsometry

Otar Bakradze*, Zurab Alimbarashvili and Rusiko Janelidze

*Institute of Cybernetics, Tbilisi, S. Euli, 5, 0186,Georgia Republic*

* Corresponding author: otobakr41@mail.ru

The paper investigates the ellipsometric method of measuring the magneto-optical parameter and optical constants in one experiment at affixed angle of the light incidence in ferromagnetics. The influence of the magnetization modulation on the change of ellipsometrical angles is shown. The method is considered for the geometries of polar, meridional and equatorial reflection Kerr effects. The formulae relating to the measured signals in polarizer-subject-analyzer (PSA) circuit with four magneto-optical and optical constants illustrated by the example of ferromagnetic Ni are given.

## Introduction

The dielectric permittivity of the magnetized crystal is a tensor of the cylindrical symmetry [1-3]

$$(\varepsilon) = \begin{vmatrix} \varepsilon & -i\varepsilon Q & 0 \\ i\varepsilon Q & \varepsilon & 0 \\ 0 & 0 & \varepsilon_0 \end{vmatrix} \quad (1)$$

where it is assumed that the magnetization is directed along the axis OZ. The magnetic permittivity at optical frequencies is taken as $(\mu) = 1$. So called magneto-optical parameter $Q = Q_1 - iQ_2$ is proportionate to the magnetization is phenomenologically relates magnetism and optics. $\varepsilon_0$ in (1) is not dependant on the magnetization, while $\varepsilon$ shows quadratic dependence what may be neglected, if $|Q| = 1$. The reflection coefficient of the non-magnetized environment is defined by Maxwell formula $\varepsilon \approx \varepsilon_0 = N^2$ where $\varepsilon = \varepsilon_1 - i\varepsilon_2$, $N = n - ik$ and $n$ is the refractive index and $k$ is the index of light absorption. The solution to the Maxwell equations with the tensor of the dielectric permittivity of kind (1) at corresponding boundary conditions gives a full description of linear magneto-optical effects. As a result of the vector value of magnetization of $\vec{M}$, they distinguish three reflection magneto-optical Kerr effects (Fig. 1). Polar and meridional effects can be defined as longitudinal, as magnetization is parallel to the plane of light incidence, while in case of equatorial or transversal effect, it is perpendicular to it. Rotation of polarization plane rotating circularly polarized waves. Ellipticity is the result of changing the reflection coefficients for the same waves. At equatorial Kerr effect, the linear component parallel to the plane of light incidence changes in amplitude and phase. Usually, the measurements of optical constants and magneto-optical parameter, included in (1), are accomplished independently resulting in a significant divergence of the results. Therefore, their simultaneous measurement in one experiment is a subject of interest. For this purpose, an ellipsometric method of calculations by using the magnetization modulation is proposed. This case may be determined as a magneto-optical modulation ellipsometry. [4] deals with the ellipsometric method to measure the parameters of massive ferromagnetics based on the equatorial Kerr effect. The same method is considered for magnetic films [5].

## Magneto-optical effects

In order to derive ellipsometric relations, we will need an explicit form of the magneto-optical Kerr effects. Under the Snell Law, $\sin\phi \approx \sin\varphi/N$, where $\varphi$ and $\phi$ are the angle of light incidence and angle of light refraction, respectively. For strongly absorbing ferromagnetics, the angle of refraction is little and therefore, an approximation $\cos\phi \approx 1$ [2, 3] is introduced. We will consider the magneto-optical effects by considering the Snell Law, i.e. without the above-mentioned approximation.

**1. Polar Kerr effect (Fig. 1a)**

The complex polar Kerr effect for $p$- and $s$-polarizations is defined as follows

$$(\chi_p)^{pol} = (\delta r_p)^{pol}/r_p \ , \ \chi_s^{pol} = -(\delta r_s)^{pol}/r_s \qquad (2)$$

where

$$r_p = (N\cos\varphi - \cos\phi)/(N\cos\varphi + \cos\phi) \ , \qquad (3)$$
$$r_s = (\cos\varphi - N\cos\phi)/(\cos\varphi + N\cos\phi)$$

are the Fresnel reflection indices. The changes of the reflection indices of the magnetized medium $(\delta r_{p,s})^{pol}$ will be [2]

$$(\delta r_p)^{pol} = (\delta r_s)^{pol} = -iQN\cos\varphi\cos^2\phi/[(N\cos\varphi + \cos\phi)(\cos\varphi + N\cos\phi)] \ . \qquad (4)$$

Let us combine the effect of both polarizations by inserting (4) in (2)

$$\chi_{p,s}^{pol} = -iQN\cos\varphi\cos^2\phi/[(\varepsilon - 1)\cos(\varphi \pm \phi)] \qquad (5)$$

where the upper sign denotes the $p$-polarization, and the lower sign denotes the $s$-polarization of light. The effect $\chi$ can be presented as a variable on the complex plane of the light polarization modes. Then the angle of rotation of the plane of polarization $\alpha$ and angle of ellipticity $\theta$ can be defined at point $\chi$ of the given complex plane

$$\tan 2\alpha = 2Re(\chi)/(1-|\chi|^2) \ , \ \sin 2\theta = 2Im(\chi)/(1+|\chi|^2) \ . \qquad (6)$$

The sign before the imaginary part $\chi$ indicates the right and left direction of the by pass of the ellipse of the reflected light. For small values, when $|\chi|=1$, we simply have $\alpha \approx Re(\chi)$, $\theta = Im(\chi)$. By applying the Snell Law, the expression (5) can be presented in the form convenient for calculations

$$\chi_{p,s}^{pol} = -iQ\cos\varphi(\varepsilon - \sin^2\varphi)/[(\varepsilon - 1)(N\cos\varphi\cos\phi \mp \sin^2\varphi)] \ . \qquad (7)$$

At a normal incidence of light $\varphi = \phi = 0$, the difference between $p$- and $s$-polarizations abolishes and the effect equals to [2]

$$\chi^{pol} = -iQN/(\varepsilon - 1) \ . \qquad (8)$$

**2. Meridional Kerr effect (Fig. 1b)**

In case of meridional geometry of the effect, the correlations (2) remain valid, and the changes in the reflection indices will equal to

$$(\delta r)^{mer} = -iQN\cos\varphi\cos\phi\sin\phi/[(N\cos\varphi + \cos\phi)(\cos\varphi + N\cos\phi)] \ . \qquad (9)$$

The meridional effect equals to

$$\chi_{p,s}^{mer} = \mp iQ\sin 2\varphi\cos\phi/[2(\varepsilon - 1)\cos(\varphi \pm \phi)] \qquad (10)$$

or

$$\chi_{p,s}^{mer} = \mp iQN\sin 2\varphi\cos\phi/[2(\varepsilon - 1)(N\cos\varphi\cos\phi \mp \sin^2\varphi)] \ . \qquad (11)$$

**3. Equatorial Kerr effect (Fig. 1c)**

The equatorial or transverse Kerr effect is essentially different from the longitudinal effects. In the first approximation by $Q$, no effect on component s is observed

$$\chi_p^{eqv} = (\delta r_p)^{eqv}/r_p \ , \ \chi_s^{eqv} = 0 \qquad (12)$$

where [3]

$$(\delta r_p)^{eqv} = -iQ\sin 2\varphi/(N\cos\varphi + \cos\phi)^2 \ . \qquad (13)$$

Then the amplitude equatorial Kerr effect will equal to
$$\chi_p^{eqv} = -iQ\varepsilon \sin 2\varphi / [\varepsilon(\varepsilon \cos^2 \varphi - 1) + \sin^2 \varphi] \ . \tag{14}$$

Unlike the longitudinal effects, no twisting of the plane of polarization is observed, but the amplitude is changed and the phase of the reflected $p$ wave is shifted. The directly measured relative light intensity will equal to [3]
$$\delta_p^{eqv} = 2Re(\chi_p^{eqv}) = 2\sin 2\varphi(Q_1 G - Q_2 H)/(G^2 + H^2) \tag{15}$$

where
$$G = \varepsilon_2[\cos^2 \varphi - \sin^2 \varphi/(\varepsilon_1^2 + \varepsilon_2^2)] \ , \tag{16}$$
$$H = \varepsilon_1[\cos^2 \varphi + \sin^2 \varphi/(\varepsilon_1^2 + \varepsilon_2^2)] - 1 \ .$$

If optical constants are known $N = n - ik$, the magneto-optical parameter $Q = Q_1 - iQ_2$ from equation (15) can be determined for two angles of the light incidence. Fig. 2 shows the dispersion of the equatorial Kerr effect in $Ni$ [6] at the angles of the light incidence of $\varphi = 70°$ and $\varphi = 80°$, and Fig. 3 shows the values $Q_1$ and $Q_2$ calculated by using these data. In the infrared spectrum $\lambda > 2mkm$ of $Ni$, these two equations become linearly dependant and no magneto-optical parameter can be determined. The ellipsometric method implies taking measurements at one fixed angle of the light incidence and is free from such a deficiency.

**Magneto-optical modulation ellipsometry**

The principal ellipsometric relationship for a non-magnetized medium is of the following form
$$\rho = \tan\psi \exp(i\Delta) = r_p/r_s \ , \tag{17}$$

where $\psi$ and $\Delta$ are the angles measured with an ellipsometer, which can be used to determine optical constants $n$ and $k$, included in the Fresnel indices. If an external impact, e.g. the magnetization reversal causes minor changes in the parameters of (17), such a situation in [7] is called the modulation ellipsometry. In case of magneto-optical modulation ellipsometry, the changes of $\psi$, $\Delta$, $r_p$ and $r_s$ can be considered as minor, if the magnetic parameter $|Q| = 1$. Then, (17) can be presented in the following form
$$\tan(\psi + \delta\psi)\exp[i(\Delta + \delta\Delta)] = (r_p + \delta r_p)(r_s + \delta r_s) \ . \tag{18}$$

By disintegrating (18) into small parameters, or by taking a logarithmic differential in both parts, we will gain
$$2\delta\psi/\sin 2\psi + i\delta\Delta = \delta r_p/r_p - \delta r_s/r_s \ . \tag{19}$$

The left part of this expression contains the measured angles $\delta\psi$ and $\delta\Delta$, and its right part is the difference of the corresponding magneto-optical effects. By equating the real and imaginary parts of (19), a magneto-optical parameter $Q$ can be determined. Below, we thoroughly deal with the proposed methods for three geometries of the magneto-optical Kerr effects. By using expression (5) for the polar effect, we gain
$$\chi_p^{pol} - \chi_s^{pol} = -iQ\sin 2\varphi \sin\varphi(\varepsilon - \sin^2\varphi)/[\varepsilon(\varepsilon\cos^2\varphi - 1) + \sin^2\varphi] \tag{20}$$

and from (10) for the meridional effect
$$\chi_p^{mer} - \chi_s^{mer} = -iQ\sin 2\varphi \cos\varphi(\varepsilon - \sin^2\varphi)/[\varepsilon(\varepsilon\cos^2\varphi - 1) + \sin^2\varphi] \ . \tag{21}$$

Note that neither expression contains the angle of refraction $\phi$. If the angle of the light incidence $\varphi = p/4$, the difference for the both longitudinal effects is the same
$$\chi_p^{long} - \chi_s^{long} = -iQ\sqrt{2}(2\varepsilon - 1)/\{2[\varepsilon(\varepsilon - 2) + 1]\} \ . \tag{22}$$

For the equatorial effect, in accordance with (12) and (14), we will have
$$2(\delta\psi)^{eqv}/\sin 2\psi + i(\delta\Delta)^{eqv} = -iQ\varepsilon\sin 2\varphi/[\varepsilon(\varepsilon\cos^2\varphi - 1) + \sin^2\varphi] \ . \tag{23}$$

Therefore, the task of determining the magneto-optical parameter $Q$ comes to the determination of $n$ and $k$, as well as $\delta\psi$ and $\delta\Delta$ of corresponding magneto-optical effects. We will use a simple (polarizer-object-analyzer (PSA)) circuit of ellipsometry [7]. Let us denote a photodetector signal by J(P, A), where P and A are azimuth angles of the polarizer and analyzer, respectively. Let us fix the polarizer angle $P = +\pi/4$ and measure the photo-signals at three angles of the analyzer $A = +p/4$, $0$, $-p/4$.

$$J_\pm(+\pi/4, \pm\pi/4) = R_s(1 \pm \sin 2\psi \cos\Delta) \quad J_0(+\pi/4, 0) = R_s(1 - \cos 2\psi) , \quad (24)$$

where $R_s = |r_s|^2$ is the intensity of the reflected light for component $s$. By excluding $R_s$ from these expressions, the ellipsometric angles can be defined

$$\psi = 1/2 \cos^{-1}[1 - 2J_0/(J_+ + J_-)] , \Delta = \cos^{-1}[(J_+ - J_-)/\sin 2\psi(J_+ + J_-)] \quad (25)$$

and then, the ellipsometric relation is defined $\rho = \tan\psi \exp(i\Delta)$. The optical constants can be determined from one of the two equivalent equations

$$N = n - ik = \sin\varphi[1 + \tan^2\varphi(1-\rho)^2/(1+\rho)^2]^{1/2} = \tan\varphi[1 - 4\rho\sin^2\varphi/(1+\rho)^2]^{1/2} . \quad (26)$$

The generalized formulae of modulation, in particular, of magneto-optical ellipsometry [7] in the given case have the following form

$$\delta J/J = \delta R_s/R_s + \beta\delta\psi + \gamma\delta\Delta \quad (27)$$

where $\delta J/J$ are the relative signals of a photodetector, $\delta R_s/R_s$ is the relative change in the light intensity for component $s$. The coefficients $\beta$ and $\gamma$ depend only on ellipsometric angles $\psi$ and $\Delta$. Let us write them out for the same modes of the polarizer and analyzer

$$\beta_\pm = 2(\tan\psi \pm \cos\Delta)/(1 \pm \sin 2\psi \cos\Delta) , \quad (28)$$

$$\gamma_\pm = \mp \sin 2\psi \sin\Delta/(1 \pm \sin 2\psi \cos\Delta) \quad \beta_0 = 4/\sin 2\psi , \quad \gamma_0 = 0.$$

By denoting $\delta J/J = S$ and excluding $\delta R_s/R_s$ from three equations (27) (for three modes of the analyzer) we gain $\delta\psi$ and $\delta\Delta$ for longitudinal (polar or meridional) magneto-optical effects

$$(\delta\psi)_{long} = 1/4 \tan\psi \{[2S_0 - (S_+ + S_-)] - (S_+ - S_-)\sin 2\psi \cos\Delta\} , \quad (29)$$

$$(\delta\Delta)_{long} = 1/4 \cot\Delta\{[2S_0 - (S_+ + S_-)](1 - \tan^2\psi) +$$
$$2(S_+ - S_-)[\tan\psi \cos\Delta - 1/(\sin 2\psi \cos\Delta)]\} ,$$

For the equatorial geometry of the effect $\delta R_s/R_s = 0$, we will have

$$(\delta\psi)_{eqv} = 1/4 \cot\psi[(S_+ + S_-) + (S_+ - S_-)\sin 2\psi \cos\Delta] , \quad (30)$$

$$(\delta\Delta)_{eqv} = 1/4 \cot\Delta$$
$$\{(S_+ + S_-)(\cot^2\psi - 1) + 2(S_+ - S_-)[\cot\psi \cos\Delta - 1/(\sin 2\psi \cos\Delta)]\} .$$

Therefore, the plan of measuring the optical constants and magneto-optical parameter can be as follows: at a fixed angle of the light incidence and polarizer azimuth, the photo-signals $J_\pm$, $J_0$, $S_\pm$ and $S_0$ are measured at a specified wave-length. Then, the ellipsometric angles $\psi$ and $\Delta$ not depending on the magnetization are defined by the formulae (25), and optical constants $n$ and $k$ are determined by (26). The magneto-optical ellipsometric angles $\delta\psi$ and $\delta\Delta$ for longitudinal geometry are defined by the formulae (29), and by (30) for equatorial geometry. Finally, the real and imaginary parts in formula (19) are equated, where (20) or (21) is used for the longitudinal effect in the right part and (23) is used for the equatorial effect. The real and imaginary parts $Q_1$ and $Q_2$ of the magneto-optical parameter are defined from these equations. Fig. 4 show the dispersion of the magneto-optical ellipsometric angles $\delta\psi$ and $\delta\Delta$ for transversal geometries in $Ni$ by using the data in Fig. 3.

**Acknowledgements**
This work was supported by the Georgian National Scientific Foundation.

**Figure captures**

Fig. 1. Polar (a), meridional (b) and equatorial (c) Kerr effects

Fig. 2. Dispersive dependences of equatorial Kerr effect in Ni at two incidence angles of light of $70°$ and $80°$.

Fig. 3. Dispersive dependences of real $Q_1$ and imaginary $Q_2$ parts of a magneto-optical parameter in Ni.

Fig. 4. Dispersive dependences of ellipsometric angles $\delta\psi$ and $\delta\Delta$ for transversal geometry in Ni.

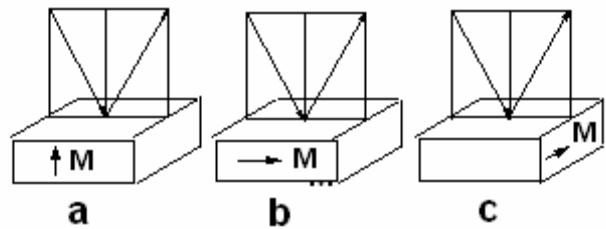

Fig.1

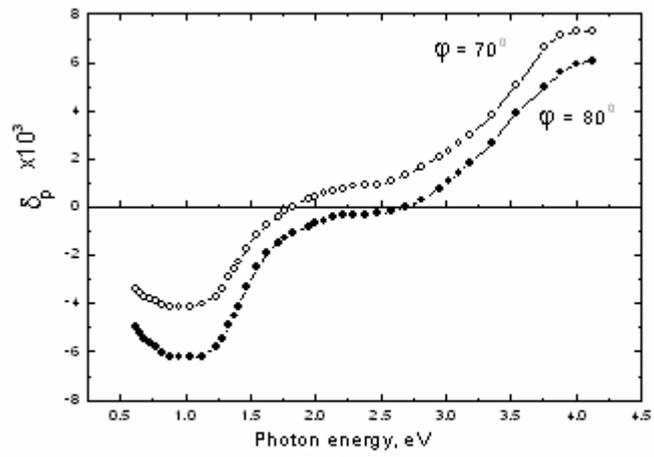
Fig. 2

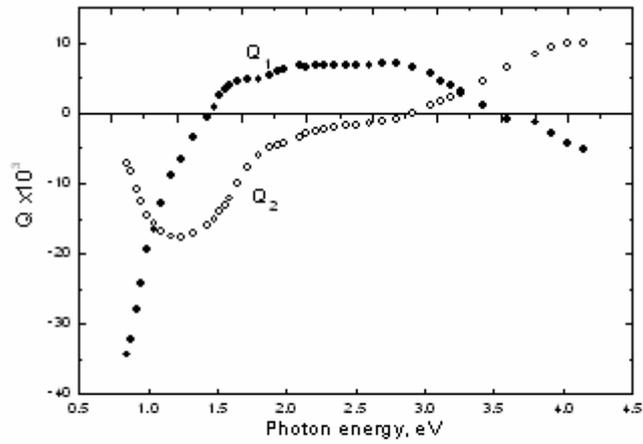
Fig. 3

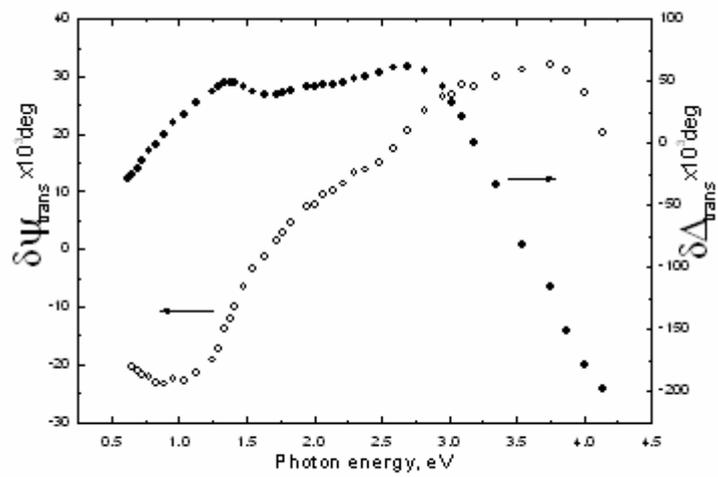
Fig. 4